\documentclass[iopfts,12pt]{iopart}
\usepackage{graphicx}
\usepackage{iopams}
\eqnobysec
\begin{document}
\title{Pseudospherical functions on a hyperboloid of one sheet}
\author{K Kowalski, J Rembieli\'nski and A Szcze\'sniak}
\address{Department of Theoretical Physics, University
of \L\'od\'z, ul.\ Pomorska 149/153,\\ 90-236 \L\'od\'z,
Poland}
\begin{abstract}
The pseudospherical functions on one-sheet, two-dimensional hyperboloid 
are discussed.  The simplest method of construction of these
functions is introduced using the Fock space structure of the
representation space of the $su(1,1)$ algebra.  The pseudospherical
functions with half-integer order are investigated.
The counterparts of the Legendre functions for the hyperboloid are 
introduced and a new class of pseudospherical functions is found.  
\end{abstract}
\vspace{1.2cm}
\section{Introduction}
The one-sheet hyperboloid is an important object in both general
relativity and quantum mechanics.  Indeed, on the one hand, it is a
model of the two-dimensional spacetime with constant curvature (de
Sitter space).  Furthermore, its invariance group $SO(2,1)$ plays
the fundamental role in the study of representations of the
$(2+1)$ - Poincar{\'e} group and corresponding relativistic wave
equations \cite{1}.  Besides of the elementary particle physics the
range of applications of the $SO(2,1)$ group include quantum optics
where this group is applied in the theory of coherent states,
especially the squeezed states \cite{2}, and the classical optics
\cite{3}.  The quantum systems related to the one-sheet hyperboloid 
were discussed in the papers [4--10].  In particular, Dane and Vardiyev 
\cite{4} studied Schr\"odinger equations with P\"oschl-Teller like
potentials connected with one and two-sheet hyperboloid and found 
the explicit expressions of the Green's functions of a free
particle on those spaces.  The contraction of eigenfunctions for
the Laplace equation on hyperboloid of one sheet and pseudoeuclidean
space have been considered in the paper of Pogosyan, Sissakian and
Winternitz \cite{5}.  The quantization of particle dynamics on
one-sheet hyperboloid embedded in three-dimensional Minkowski space
was discussed by Piechocki and Jorjadze in papers \cite{6,7,8}.  The
coherent state quantization of particle in the two-dimensional de Sitter 
space i.e.\ one-sheet hyperboloid, was investigated by Gazeau and
Piechocki in \cite{9}.  

The pseudospherical functions in the general case 
of hyperboloids of arbitrary dimensions were introduced independently, 
using different methods by 
R\hbox{\kern.1em\hbox{\char24} \kern-.9em a}czka, Limi\'c and
Niederle \cite{11} and Strichartz \cite{12}.  Both approaches are based 
on the very general mathematical scheme involving multidimensional hyperboloids,
no wonder that the physically interesting case of the
two-dimensional hyperboloid of one sheet was not discussed therein
in a more detail.  Recently, the wavefunctions for the
two-dimensional hyperboloids involving their different parametrizations
were studied in \cite{4}.  The approach used in \cite{4} for the
identification of wavefunctions in the hyperbolic parametrization of 
hyperboloids coincide with that adopted in \cite{11}.

In this work we construct the pseudospherical functions for
the quantum mechanics on a one-sheet hyperboloid using the simplest
algorithm developed for the case of the sphere $S^2$, based on
application of the Fock space structure of representations of the
$SO(2,1)$ group.  The approach taken up in this work is more general
than alternative ones mentioned earlier because it enables fixing
the phase of the pseudospherical functions.  Surprisingly, to our knowledge, 
despite the fact that the pseudospherical functions for hyperboloids are 
also referred to as the ``spherical functions'' or ``spherical harmonics'' 
(see for example \cite{13}) such natural approach was not taken up so far.
Utilizing the introduced formalism we study for the first time in
the literature the pseudospherical functions with half-integer
order.  We also identify the pseudospherical counterparts of
the Legendre functions and study their basic properties in the case
of the discrete series representations.  Finally, we find a new class 
of pseudospherical functions related to the discrete series.
\section{Basic representations}
We first discuss the basic properties of representations of the
$SO(2,1)$ group which is the invariance group for the one-sheet
hyperboloid such that
\begin{equation}
(x^1)^2+(x^2)^2-(x^3)^2=a^2,
\end{equation}
where $a>0$ is a parameter.  The generators $K_i$, $i=1,\,2,\,3$,
of the representations of $SO(2,1)$ satisfy the following
commutation relations:
\begin{equation}
[K_3,K_\alpha]={\rm i}\varepsilon_{\alpha\beta}K_\beta,
\qquad [K_\alpha,K_\beta]=-{\rm i}\varepsilon_{\alpha\beta}K_3,\qquad 
\alpha,\beta =1,\,2.
\end{equation}
The algebra $so(2,1)$ (2.2) expressed with the help of operators
$K_3$, $K_\pm=K_1\pm {\rm i}K_2$ takes the form
\begin{equation}
[K_3,K_\pm]=\pm K_\pm,\qquad [K_+,K_-]=-2K_3.
\end{equation}
The Casimir operator of the algebra (2.3) is given by
\begin{equation}
C_2=K_3^2-K_1^2-K_2^2=K_3(K_3+1)-K_-K_+=K_3(K_3-1)-K_+K_-.
\end{equation}
Consider the unitary irreducible representation of the algebra (2.3)
spanned by the (normalized) common eigenvectors $|k,m\rangle$ of the
operators $C_2$ and $K_3$:
\begin{equation}
C_2|k,m\rangle = k(k+1)|k,m\rangle,\qquad K_3|k,m\rangle = m|k,m\rangle.
\end{equation}
Recall that \cite{14}
\numparts
\begin{eqnarray}
K_+|k,m\rangle &=& \sqrt{(m-k)(m+k+1)}|k,m+1\rangle,\\
K_-|k,m\rangle &=& \sqrt{(m+k)(m-k-1)}|k,m-1\rangle.
\end{eqnarray}
\endnumparts
The unitary irreducible representations can be grouped into three
classes accordingly to the spectrum of $C_2$ and $K_3$.  The
discrete series $D_+$ and $D_-$.  The series $D_+$ is specified by
\begin{equation}
k=-\frac{1}{2},\,0,\,\frac{1}{2},\,1,\ldots ,\qquad m\ge k+1.
\end{equation}
This series is bounded below.  Namely
\begin{equation}
K_-|k,k+1\rangle = 0.
\end{equation}
Using (2.6a) we find that the vectors $|k,m\rangle$ can be obtained
from the lowest-weight state $|k,k+1\rangle$ via
\begin{equation}
|k,m\rangle = \sqrt{\frac{(2k+1)!}{(m-k-1)!(m+k)!}}K_+^{m-k-1}
|k,k+1\rangle,\qquad m\ge k+1.
\end{equation}
For the discrete series $D_-$ we have
\begin{equation}
k=-\frac{1}{2},\,0,\,\frac{1}{2},\,1,\ldots ,\qquad m\le -(k+1).
\end{equation}
The series $D_-$is bounded above.  Namely
\begin{equation}
K_+|k,-(k+1)\rangle = 0.
\end{equation}
The counterpart of (2.9) is
\begin{equation}
\fl |k,m\rangle = \sqrt{\frac{(2k+1)!}{[-(m+k+1)]!(k-m)!}}K_-^{-(m+k+1)}
|k,-(k+1)\rangle,\qquad m\le -(k+1).
\end{equation}
The second class is the continuous principal series such that
\begin{equation}
k=-\frac{1}{2}+{\rm i}\lambda,\quad \lambda >0,\quad
m=0,\,\pm1,\,\pm2,\ldots\,\,\hbox{{\rm or
}}m=\pm\frac{1}{2},\,\pm\frac{3}{2},\ldots .
\end{equation}
Finally, we have the continuous supplementary series
\begin{equation}
-\frac{1}{2}<k<0,\qquad m=0,\,\pm1,\,\pm2,\ldots.
\end{equation}
Bearing in mind the possible applications we remark that the discrete
series representations of $SU(1,1)$ which is locally isomorphic to
$SO(2,1)$, were related in \cite{15} to bound states and the
continuous series representations to the scattering states.

We now discuss the realization of generators by means of the
differential operators.  We first write down the following realization 
of the algebra (2.2):
\begin{eqnarray}
K_1&=& -{\rm i}(x^2\partial_3 + x^3\partial_2),\nonumber\\
K_2&=& {\rm i}(x^1\partial_3 + x^3\partial_1),\\
K_3&=& -{\rm i}(x^1\partial_2 - x^2\partial_1),\nonumber
\end{eqnarray}
where $\partial_i\equiv\partial/\partial x^i$, $i=1,\,2,\,3$.
Introducing the polar hyperbolic coordinates (biharmonic
coordinates) such that
\begin{eqnarray}
x^1&=&a\cosh\tau\cos\varphi,\nonumber\\
x^2&=&a\cosh\tau\sin\varphi,\\
x^3&=&a\sinh\tau,\nonumber
\end{eqnarray}
so the hyperboloid (2.1) is specified by $a=const$, $\tau\in{\mathbb
R}$, and $\varphi\in[0,2\pi)$, we get from (2.15)
\numparts
\begin{eqnarray}
K_+&=&-e^{\rm i\varphi}\left(\frac{\partial}{\partial\tau}+\rm i
{\rm tgh}\tau\frac{\partial}{\partial\varphi}\right),\\
K_-&=&e^{-\rm i\varphi}\left(\frac{\partial}{\partial\tau}-\rm i
{\rm tgh}\tau\frac{\partial}{\partial\varphi}\right),\\
K_3&=&-{\rm i}\frac{\partial}{\partial\varphi}.
\end{eqnarray}
\endnumparts
Hence
\begin{equation}
C_2=\frac{\partial^2}{\partial\tau^2}+{\rm tgh}\tau\frac{\partial}{\partial\tau}
-\frac{1}{\cosh^2\tau}\frac{\partial^2}{\partial\varphi^2}.
\end{equation}
Clearly, the pseudospherical functions ${\cal Y}^m_k(\tau,\varphi)$, i.e.\ the
realizations of the abstract vectors $|k,m\rangle$ in the coordinate
representations, satisfy the system of differential equations
\numparts
\begin{eqnarray}
C_2{\cal Y}^m_k(\tau,\varphi)&=&k(k+1){\cal Y}^m_k(\tau,\varphi),\\
K_3{\cal Y}^m_k(\tau,\varphi)&=&m{\cal Y}^m_k(\tau,\varphi),
\end{eqnarray}
\endnumparts
where $C_2$ and $K_3$ are given by (2.18) and (2.17c), respectively.
Separating variables we find that ${\cal Y}^m_k(\tau,\varphi)$ is of the form
\begin{equation}
{\cal Y}^m_k(\tau,\varphi)=e^{{\rm i}m\varphi}f^m_k(\tau),
\end{equation}
where $f^m_k(\tau)$ fulfils
\begin{equation}
\frac{d^2f^m_k(\tau)}{d\tau^2}+{\rm tgh}\tau\frac{df^m_k(\tau)}{d\tau}
+\left[-k(k+1)+\frac{m^2}{\cosh^2\tau}\right]f^m_k(\tau)=0.
\end{equation}
In other words, the system (2.19) can be brought down to eq.\ (2.21).
\section{Pseudospherical functions for discrete series representations}
We now discuss the pseudospherical functions ${\cal
Y}^m_k(\tau,\varphi)$ in the case of the discrete series $D_+$.
The equation (2.21) was the point of departure for finding pseudospherical 
functions in \cite{11}.  The solution to (2.21) was actually guessed
in \cite{11} (see (3.17) below).  We now apply the simpler and most natural method for 
constructing the pseudospherical functions based on the well-known approach 
taken up for spherical functions (see for example \cite{16}).  Consider the case 
of the series $D_+$.  By the last equation of (2.4) the wavefunction
\begin{equation}
{\cal Y}^{k+1}_k(\tau,\varphi)=e^{{\rm i}(k+1)\varphi}f^{k+1}_k(\tau),
\end{equation}
representing the lowest-weight state $|k,k+1\rangle$, satisfying
\begin{equation}
K_-{\cal Y}^{k+1}_k(\tau,\varphi)=0,
\end{equation}
is the solution of the system (2.19) with $m=k+1$.  Using (2.17b) we
get from (3.2) the first order equation
\begin{equation}
\frac{df^{k+1}_k(\tau)}{d\tau} + (k+1){\rm tgh}\tau f^{k+1}_k(\tau)=0.
\end{equation}
The solution of the elementary equation (3.3) is
\begin{equation}
f^{k+1}_k(\tau)=c_k\cosh^{-(k+1)}\tau,
\end{equation}
where $c_k$ is a normalization constant.  Now, the Hilbert space of
square integrable functions on a hyperboloid of one sheet (2.1) is
specified by the scalar product
\begin{equation}
\langle F|G\rangle = \int_{-\infty}^{\infty}\cosh\tau d\tau
\int_{0}^{2\pi}d\varphi F^*(\tau,\varphi)G(\tau,\varphi),
\end{equation}
where we set without loose of generality $a=1$.  Hence, using the
identity \cite{17}
\begin{equation}
\int_{0}^{\infty}\frac{dx}{\cosh^{2k+1}x}=\frac{(2k-1)!!}{(2k)!!}
\frac{\pi}{2},
\end{equation}
where $k=0,\,1,\,2,\ldots,$ we get
\begin{equation}
c_k=\frac{1}{\sqrt{2}\pi}\sqrt{\frac{(2k)!!}{(2k-1)!!}}=\frac{2^{k-
\frac{1}{2}}}{\pi}\frac{k!}{\sqrt{(2k)!}}.
\end{equation}
Furthermore, taking into account (2.17a) we obtain the identity
\begin{equation}
K_+^r[e^{{\rm i}m\varphi}f(\tau)] = (-1)^re^{{\rm i}(m+r)\varphi}\left(
\cosh^{m+r}\tau\frac{d^r}{d(\sinh\tau)^r}\cosh^{-m}\tau\right)f(\tau).
\end{equation}
An immediate consequence of (2.9), (3.8), (3.1), (3.4) and (3.7)
is the following formula on the pseudospherical functions ${\cal Y}^m_k$:
\begin{eqnarray}
\fl {\cal Y}^m_k(\tau,\varphi)=(-1)^{m-k-1}2^kk!\sqrt{\frac{2k+1}{2\pi^2(m-k-1)!(m+k)!}}
e^{{\rm i}m\varphi}\cosh^m\tau\frac{d^{m-k-1}}{d(\sinh\tau)^{m-k-1}}
\cosh^{-2(k+1)}\tau\nonumber\\
\end{eqnarray}
The relation (3.9) can be written as
\begin{equation}
{\cal Y}^m_k(\tau,\varphi)=\sqrt{\frac{2k+1}{2\pi^2(m-k-1)!(m+k)!}}e^{{\rm i}m\varphi}
{\cal P}^m_k(\sinh\tau),
\end{equation}
where ${\cal P}^m_k(x)$ are counterparts of the (associated)
Legendre functions connected with spherical harmonics, defined by
\begin{equation}
{\cal P}^m_k(x)=(-1)^{m-k-1}2^kk!(1+x^2)^{\frac{m}{2}}\frac{d^{m-k-1}}
{dx^{m-k-1}}\frac{1}{(1+x^2)^{k+1}}.
\end{equation}
The basic recurrences satisfied by the functions ${\cal P}^m_k$
analogous to the well-known formulas on the associated Legendre
polynomials are introduced in Appendix.  Now taking into account the
identity \cite{18}
\begin{equation}
\frac{d^n}{dx^n}f(x^2)=n!\sum_{r=0}^{[\frac{n}{2}]}\frac{(2x)^{n-2r}}{r!
(n-2r)!}f^{(n-r)}(x^2),
\end{equation}
where $[p]$ is the biggest integer in $p$, we find
\begin{equation}
\fl {\cal
P}^m_k(x)=2^k(m-k-1)!(1+x^2)^{\frac{m}{2}}\sum_{r=0}^{[\frac{m-k-1}{2}]}
(-1)^r\frac{(m-r-1)!}{r!(m-k-1-2r)!}\frac{(2x)^{m-k-1-2r}}{(1+x^2)^{m-r}}
\end{equation}
which leads to
\begin{eqnarray}
\fl {\cal
P}^m_k(\sinh\tau)=2^k(m-k-1)!\cosh^{-(k+1)}\tau\sum_{r=0}^{[\frac{m-k-1}{2}]}
(-1)^r\frac{(m-r-1)!}{r!(m-k-1-2r)!}(2{\rm tgh}\tau)^{m-k-1-2r}\nonumber\\
\end{eqnarray}
Hence, using (3.10) the pseudospherical functions can be written in explicit
form as
$$\displaylines{
{{\cal Y}^m_k(\tau,\varphi)}=2^k\sqrt{\frac{(2k+1)(m-k-1)!}{2\pi^2(m+k)!}}
e^{{\rm i}m\varphi}\cosh^{-(k+1)}\tau\hfill\cr
\qquad\qquad\quad\times\sum_{r=0}^{[\frac{m-k-1}{2}]}(-1)^r\frac{(m-r-1)!}
{r!(m-k-1-2r)!}(2{\rm tgh}\tau)^{m-k-1-2r}.\hfill\llap(3.15)\cr}$$
The authors did not find the formula on the pseudospherical functions of the
form (3.15) in the literature.  The sum from (3.15) and thus 
${\cal Y}^m_k(\tau,\varphi)$ can be expressed in terms of the
hypergeometric functions ${}_2F_1(\alpha,\beta;\gamma;z)$.  Namely,
we find after some calculations
$$\displaylines{
{\cal Y}^m_k(\tau,\varphi)=(-1)^{\frac{1}{2}(m-k-1)}\sqrt{
\frac{(2k+1)\Gamma(\hbox{$\scriptstyle{1\over 2}$}(m-k))\Gamma(
\hbox{$\scriptstyle{1\over 2}$}(m+k+1))}
{4\pi^2\Gamma(\hbox{$\scriptstyle{1\over 2}$}
(m-k+1))\Gamma(\hbox{$\scriptstyle{1\over 2}$}(m+k+2))}}e^{{\rm
i}m\varphi}\cosh^{-(k+1)}\tau\hfill\cr
\qquad\qquad\quad\times\,{}_2F_1[\hbox{$\scriptstyle{1\over 2}$}(m+k+1),
\hbox{$\scriptstyle{1\over 2}$}(-m+k+1)
;\hbox{$\scriptstyle{1\over 2}$};{\rm tgh}^2\tau],\hfill\llap(3.16a)\cr}$$
where $m-k=2n+1$, $n=0,\,1,\,2,\,\ldots$, $\Gamma(x)$ is the gamma function, and
$$\displaylines{
{\cal Y}^m_k(\tau,\varphi)=(-1)^{\frac{1}{2}(m-k-2)}\sqrt{
\frac{(2k+1)\Gamma(\hbox{$\scriptstyle{1\over 2}$}(m-k+1))\Gamma(
\hbox{$\scriptstyle{1\over 2}$}(m+k+2))}
{\pi^2\Gamma(\hbox{$\scriptstyle{1\over 2}$}
(m-k))\Gamma(\hbox{$\scriptstyle{1\over 2}$}(m+k+1))}}e^{{\rm
i}m\varphi}\cosh^{-(k+1)}\tau{\rm tgh}\tau\hfill\cr
\qquad\qquad\quad\times\,{}_2F_1[\hbox{$\scriptstyle{1\over 2}$}(m+k+2),
\hbox{$\scriptstyle{1\over 2}$}(-m+k+2)
;\hbox{$\scriptstyle{3\over 2}$};{\rm tgh}^2\tau],\hfill\llap(3.16b)\cr}$$
where $m-k=2n+2$ and $n=0,\,1,\,2,\,\ldots$.  Up to the absent phase factor 
$(-1)^{\frac{1}{2}(m-k-1)}$ in (3.16a) and the phase factor 
$(-1)^{\frac{1}{2}(m-k-2)}$ in (3.16b) as well as some typo in the formula
referring to (3.16b) the solution (3.16) was guessed from (2.21) by 
R\hbox{\kern.1em\hbox{\char24} \kern-.9em a}czka, Limi\'c and
Niederle \cite{11} and rediscovered by Dane and Verdiyev \cite{4}.
More precisely, one can put
\setcounter{equation}{16}%
\begin{equation}
f^m_k(\tau) = \cosh^\alpha\tau g^m_k(\tau)
\end{equation}
which leads to
\begin{equation}
\fl\frac{d^2g^m_k(\tau)}{d\tau^2}+(2\alpha+1){\rm tgh}\tau\frac{dg^m_k(\tau)}{d\tau}
+\left[\alpha(\alpha+1)-k(k+1)+\frac{m^2-\alpha^2}{\cosh^2\tau}\right]g^m_k(\tau)=0.
\end{equation}
Hence setting $\alpha=-(k+1)$ and making the ansatz
\begin{equation}
g^m_k(\tau) = h^m_k({\rm tgh}^2\tau)
\end{equation}
we obtain the hypergeometric equation
\begin{equation}
\fl x(1-x){h^m_k}{}''(x)+[\hbox{$\scriptstyle{1\over 2}$}-(k+2)x]h^m_k{}'(x)+\hbox{$\scriptstyle
{1\over 4}$}[m^2-(k+1)^2]h^m_k(x)=0
\end{equation}
satisfied by the hypergeometric function ${}_2F_1[\hbox{$\scriptstyle{1\over 2}$}(m+k+1),
\hbox{$\scriptstyle{1\over 2}$}(-m+k+1);\hbox{$\scriptstyle{1\over
2}$};x]$.  As is well known if ${}_2F_1(\alpha,\beta;\gamma;z)$ is a
solution of the hypergeometric equation and $\gamma$ is not integer,
then the second linearly independent solution is
$z^{1-\gamma}{}_2F_1(\alpha-\gamma +1,\beta-\gamma+1;2-\gamma;z)$.
Therefore, the second linearly independent solution to (3.20) is
$\sqrt{x}{}_2F_1[\hbox{$\scriptstyle{1\over 2}$}(m+k+2),
\hbox{$\scriptstyle{1\over 2}$}(-m+k+2);\hbox{$\scriptstyle{3\over
2}$};x]$.  On putting in the both solutions $x={\rm tgh}^2\tau$ and
using (3.17), (3.19) and (2.20) we arrive, up to the normalization 
constant, at the solutions (3.16a) and (3.16b), respectively.  We
point out that the problem of normalization of solutions to (3.20)
is a nontrivial task based on utilization of some identities
satisfied by hypergeometric functions (compare \cite{4}).  On the other 
hand, the normalization of pseudospherical functions in our approach
is ensured by normalization of the elementary function (3.4).

We stress that the phase factor in (3.16a) and (3.16b) is not
arbitrary and its correct form ensures that $K_+$ and $K_-$ act on
functions ${\cal Y}^m_k(\tau,\varphi)$ as ladder operators in
accordance with (2.6).  Clearly, the phase cannot be fixed based
only on eq.\ (2.21).

The possibility of generation of pseudospherical functions by the
action of the raising operator on the lowest-weight state and lowering
operator on the highest-weight state in the case of the series $D_+$
and $D_-$, respectively was recognized in \cite{1}. However, no explicit
formulae on the pseudospherical functions in the general case of
arbitrary $m\ge k+1$ (series $D_+$), and $m\le-(k+1)$ (series $D_-$)
were provided in \cite{1}.

As remarked in \cite{1} by making the ansatz
\begin{equation}
f^m_k(\tau) = u^m_k({\rm i}\sinh\tau)
\end{equation}
one can reduce (2.21) to the equation of the form
\begin{equation}
(1-z^2)u^m_k{}''(z) -
2zu^m_k{}'(z)+\left[k(k+1)-\frac{m^2}{1-z^2}\right]u^m_k(z)=0
\end{equation}
satisfied by the associated Legendre functions $P^m_k(z)$ (spherical
harmonics). Nevertheless, it is not clear what is advantage of such
reduction.  For example, we have $P^m_k(z)\equiv0$ for $m\ge k+1$ \cite{21}, 
while this inequality is satisfied by indices of all pseudospherical 
functions in the case of the representation $D_+$.  Furthermore, the
divergence of the norm of the functions $P^m_k({\rm i}\sinh\tau)$
in the case of the continuous principal series with
$k=-\frac{1}{2}+{\rm i}\lambda$ was reported in \cite{1}.

We finally point out that the counterpart of the relations (3.16) in the 
case of the spherical harmonics is of the form
$$\displaylines{
Y^m_j(\theta,\varphi)=a_{mj}e^{{\rm i}m\varphi}\sin^j\theta\,{}_2F_1(
-\hbox{$\scriptstyle{1\over 2}$}(j-m),
-\hbox{$\scriptstyle{1\over 2}$}(j+m)
;\hbox{$\scriptstyle{1\over 2}$};-{\rm ctg}^2\theta),\hfill\llap(3.23a)\cr}$$
where $j-m=2n$, and $n=1,\ldots ,j$; $a_{mj}$ is a normalization
constant, and
$$\displaylines{
Y^m_j(\theta,\varphi)=b_{mj}e^{{\rm i}m\varphi}\sin^j\theta{\rm ctg}\theta\,{}_2F_1(
-\hbox{$\scriptstyle{1\over 2}$}(j-m-1),
-\hbox{$\scriptstyle{1\over 2}$}(j+m-1)
;\hbox{$\scriptstyle{3\over 2}$};-{\rm ctg}^2\theta),\hfill\llap(3.23b)\cr}$$
where $j-m=2n-1$, $n=1,\ldots ,j$, and $b_{mj}$ is a normalization
constant.

We now discuss the pseudospherical functions in the case of the series $D_{-}$.
Taking into account the second equation of (2.4) we find that the
wavefunction
\setcounter{equation}{23}%
\begin{equation}
\tilde{\cal Y}^{-(k+1)}_k(\tau,\varphi)=e^{-{\rm
i}(k+1)\varphi}f^{-(k+1)}_k(\tau)
\end{equation}
representing the highest-weight state $|k,-(k+1)\rangle$, such that
\begin{equation}
K_+\tilde{\cal Y}^{-(k+1)}_k(\tau,\varphi)=0,
\end{equation}
is the solution to (3.5), where $m=-(k+1)$.  Moreover, it follows
immediately from (3.3a) and (3.25) that the function
$f^{-(k+1)}_k(\tau)$ fulfils the same equation (3.3) as
$f^{k+1}_k(\tau)$, so we set $f^{-(k+1)}_k(\tau)=f^{k+1}_k(\tau)$.
Furthermore, proceeding analogously as with (3.8) we get
\begin{equation}
K_-^r[e^{{\rm i}m\varphi}f(\tau)] = e^{{\rm i}(m-r)\varphi}\left(
\cosh^{-m+r}\tau\frac{d^r}{d(\sinh\tau)^r}\cosh^m\tau\right)f(\tau).
\end{equation}
Eqs.\ (2.12), (3.26), (3.4) and (3.7) taken together yield
\begin{eqnarray}
\fl \tilde{\cal
Y}^m_k(\tau,\varphi)=2^kk!\sqrt{\frac{2k+1}{2\pi^2[-(m+k+1)]!(k-m)!}}
e^{{\rm i}m\varphi}\cosh^{-m}\tau\frac{d^{-(m+k+1)}}{d(\sinh\tau)^{-(m+k+1)}}
\cosh^{-2(k+1)}\tau\nonumber\\
\end{eqnarray}
The pseudospherical functions $\tilde{\cal Y}^m_k(\tau,\varphi)$ can be
expressed by means of the functions ${\cal P}^m_k(\sinh\tau)$ defined by
(3.11) in the form
\begin{equation}
\fl \tilde{\cal
Y}^m_k(\tau,\varphi)=(-1)^{m+k+1}\sqrt{\frac{2k+1}{2\pi^2[-(m+k+1)]!(k-m)!}}e^{{\rm
i}m\varphi}
{\cal P}^{-m}_k(\sinh\tau).
\end{equation}
Comparing (3.28) and (3.10) we obtain the following relationship
between the pseudospherical functions ${\cal Y}^m_k$ and $\tilde{\cal Y}^m_k$
referring to the discrete series $D_+$ and $D_-$, respectively:
\begin{equation}
[{\cal Y}^m_k(\tau,\varphi)]^* = (-1)^{m-k-1}\tilde{\cal Y}^{-m}_k(\tau,\varphi).
\end{equation}

For the sake of completeness we finally write down the
Laplace-Beltrami operator in the biharmonic coordinates (2.16)
\begin{eqnarray}
\fl\Delta &=& \left(\frac{\partial}{\partial x^1}\right)^2+
\left(\frac{\partial}{\partial x^2}\right)^2-
\left(\frac{\partial}{\partial x^3}\right)^2=
\frac{1}{a}\frac{\partial^2}{\partial a^2}a
-\frac{1}{a^2}\left(\frac{\partial^2}{\partial\tau^2}+
{\rm tgh}\tau\frac{\partial}{\partial\tau}
-\frac{1}{\cosh^2\tau}\frac{\partial^2}{\partial\varphi^2}\right)\nonumber\\
\fl {}&=& \frac{1}{a}\frac{\partial^2}{\partial a^2}a-\frac{1}{a^2}C_2,
\end{eqnarray}
where $C_2$ is the Casimir operator given by (2.18).  It follows
directly from (3.30) and (2.19a) that the functions defined by
\begin{equation}
p^m_k(a) = a^k{\cal Y}^m_k(\tau,\varphi),
\end{equation}
fulfil
\begin{equation}
\Delta p^m_k(a) = 0.
\end{equation}
The functions $p^m_k(a)$ are hyperbolic counterparts of the harmonic
polynomials also called volume spherical functions.
\section{A new class of pseudospherical functions}
We now return to (2.21).  Motivated by the a priori possible joining
of the series $D_+$ and $D_-$ we relax from the requirement $m\ge
k+1$ for the series $D_+$ and we study the case $m=k$ in (2.21).  We
then have
\begin{equation}
\frac{d^2f^k_k(\tau)}{d\tau^2}+{\rm tgh}\tau\frac{df^k_k(\tau)}{d\tau}
-(k^2{\rm tgh}^2\tau+k)f^k_k(\tau)=0.
\end{equation}
We point out that the same equation is obtained from (2.21) for $m=-k$.
Hence we get $f^{-k}_k(\tau)=f^k_k(\tau)$.  On introducing the new 
functions $g_k$ such that
\begin{equation}
f^k_k(\tau)=g_k({\rm tgh}\tau),
\end{equation}
we arrive at the equation
\begin{equation}
g''_k(x)-\frac{x}{1-x^2}g'_k(x)-\frac{k^2x^2+k}{(1-x^2)^2}g_k(x)=0.
\end{equation}
The general solution to (4.3) is given by (see \cite{19}, eq.\ (2.76a))
\begin{equation}
g_k(x) = (1-x^2)^{-\frac{k}{2}}[\alpha_k
+\beta_k\int(1-x^2)^{k-\frac{1}{2}}dx], 
\end{equation}
where $\alpha_k$ and $\beta_k$ are integration constants.  From (4.4) we
immediately get
\numparts
\begin{eqnarray}
g_0(x) &=& \alpha_0 +\beta_0\arcsin x,\\
g_1(x) &=& (1-x^2)^{-\frac{1}{2}}\{\alpha_1
+\hbox{$\scriptstyle\beta_1\over2$}[x(1-x^2)^{\frac{1}{2}}+\arcsin x]\}.
\end{eqnarray}
\endnumparts
The remaining functions $g_k(x)$, $k\ge2$, can be obtained from
(4.4) and the following identity \cite{17}:
$$\displaylines{
\int(1-x^2)^{k-\frac{1}{2}}dx=\frac{x(1-x^2)^\frac{1}{2}}{2k}
\left[\phantom{\frac{1}{2}}\!\!\!(1-x^2)^{k-1}\right.\hfill\cr
+ \left.\sum_{r=1}^{k-1}\frac{(2k-1)(2k-3)\cdots
[2k-(2r-1)]}{2^r(k-1)(k-2)\cdots(k-r)}(1-x^2)^{k-1-r}\right]
+\frac{(2k-1)!!}{2^kk!}\arcsin x,\hfill\llap(4.6a)\cr}$$
which can be written in a simpler form
$$\displaylines{
\int(1-x^2)^{k-\frac{1}{2}}dx=\frac{x(1-x^2)^\frac{1}{2}}{2k}
\left[\phantom{\frac{1}{2}}\!\!\!(1-x^2)^{k-1}\right.\hfill\cr
{}+ \left.\frac{(2k-1)!!}{2^k(k-1)!(1-x^2)}\sum_{r=1}^{k-1}
\frac{2^r(r-1)!}{(2r-1)!!}(1-x^2)^r\right]
+\frac{(2k-1)!!}{2^kk!}\arcsin x,\qquad k\ge2.\hfill\llap(4.6b)\cr}$$
Now eqs.\ (2.20), (4.2), (4.4) and (4.6) taken together yield
$$\displaylines{
{\cal Y}^0_0(\tau,\varphi) = \alpha_0 +\beta_0\arcsin{\rm tgh}\tau
\hfill\llap(4.7a)\cr
{\cal Y}^1_1(\tau,\varphi)=\alpha_1 e^{{\rm i}\varphi}\cosh\tau
+\frac{\beta_1}{2}e^{{\rm i}\varphi}({\rm
tgh}\tau+\cosh\tau\arcsin{\rm tgh}\tau)\hfill\llap(4.7b)\cr
{\cal Y}^k_k(\tau)=\alpha_k e^{{\rm i}k\varphi}\cosh^k\tau +
\beta_k e^{{\rm i}k\varphi}\cosh^k\tau\left\{\frac{{\rm
tgh}\tau}{2k\cosh\tau}\left[\phantom{\frac{1}{2}}\!\!\!\cosh^{-2k+2}\tau\right.
\right.\hfill\cr
{}+\frac{(2k-1)!!\cosh^2\tau}{2^k(k-1)!}\sum_{r=1}^{k-1}\left.\left.\frac{2^r(r-1)!}
{(2r-1)!!}\cosh^{-2r}\tau\right] +\frac{(2k-1)!!}{2^kk!}\arcsin{\rm tgh}\tau
\right\},\quad k\ge 2,\hfill\llap(4.7c)\cr
}$$
where the use can be made of the identity
\setcounter{equation}{7}%
\begin{equation}
\arcsin{\rm tgh}\tau = 2{\rm arctg} e^\tau-\hbox{$\scriptstyle\pi\over2$}.
\end{equation}
As far as we are aware the formulas (4.7) on the pseudospherical
functions are new.  The functions ${ \cal Y}^k_k(\tau,\varphi)$, 
$k=0,\,1,\,2,\,\ldots\,$, and
\begin{equation}
\tilde{\cal Y}^{-k}_k(\tau,\varphi)=e^{-{\rm i}k\varphi}f^{-k}_k(\tau)=e^{-{\rm
i}k\varphi}f^k_k(\tau)
\end{equation}
are related to the series $D_-$ and $D_+$, respectively.  Namely, we
have
\begin{equation}
K_-^{2k+1}{\cal Y}^k_k\sim \tilde{\cal Y}^{-(k+1)}_k,\qquad
K_+^{2k+1}\tilde{\cal Y}^{-k}_k\sim {\cal Y}^{k+1}_k.
\end{equation}
Notice that by virtue of (4.10), in the case of $k=0$, the function
${\cal Y}^0_0$ really joins both series $D_+$ and $D_-$ via the ladder
operators $K_+$ and $K_-$.  Finally,
we remark that the functions ${\cal Y}^k_k(\tau,\varphi)$ are not square
integrable on the one-sheet hyperboloid.  It should also be noted
that the function ${\cal Y}^0_0$ for $\beta_0=0$ is constant and
refers to the trivial representation.
\section{Pseudospherical functions for continuous principal series}
\subsection{Normalization of pseudospherical functions}
In this section we study the pseudospherical functions for
continuous principal series.  In the case of such series we have
neither highest nor lowest weight states and therefore the spherical
functions cannot be found by solving any first-order differential
equation as for the discrete series and application of the ladder
operators.  As a consequence, we have two kinds of pseudospherical
functions referring to two linearly independent solutions of the
second order equation (2.21).  In view of (3.16) and the form of the
parameter $k$ in the case of the continuous principal series such that
$k=-\frac{1}{2}+{\rm i}\lambda$, these functions can be written as
\begin{eqnarray}
\fl{\hat{\cal Y}}^m_\lambda(\tau,\varphi)=e^{{\rm i}m\varphi}{\hat c}_{m\lambda}
\cosh^{-(\frac{1}{2}+{\rm i}\lambda)}\tau{}_2F_1[\hbox{$\scriptstyle{1\over 2}$}(m+
\hbox{$\scriptstyle{1\over 2}$}+{\rm i}\lambda),
\hbox{$\scriptstyle{1\over 2}$}(-m+\hbox{$\scriptstyle{1\over 2}$}+{\rm i}\lambda)
;\hbox{$\scriptstyle{1\over 2}$};{\rm tgh}^2\tau],\\
\fl{\bar{\cal Y}}^m_\lambda(\tau,\varphi)=e^{{\rm i}m\varphi}{\bar c}_{m\lambda}
\cosh^{-(\frac{1}{2}+{\rm i}\lambda)}\tau{\rm tgh}\tau
{}_2F_1[\hbox{$\scriptstyle{1\over 2}$}
(m+\hbox{$\scriptstyle{3\over 2}$}+{\rm i}\lambda),
\hbox{$\scriptstyle{1\over 2}$}(-m+\hbox{$\scriptstyle{3\over 2}$}+{\rm i}\lambda)
;\hbox{$\scriptstyle{3\over 2}$};{\rm tgh}^2\tau],
\end{eqnarray}
where ${\hat c}_{m\lambda}$ and ${\bar c}_{m\lambda}$ are
normalization constants.  Following the algoritm described in
\cite{4} we now normalize the function ${\hat {\cal Y}}^m_\lambda
(\tau,\varphi)$.  Firstly, using the formula on the analytic
continuation \cite{20} such that
\begin{eqnarray}
&&\fl {}_2F_1(\alpha,\beta ;\gamma;z) = A_1\,{}_2F_1(\alpha,\beta
;\alpha+\beta-\gamma+1;1-z)\nonumber\\
&&\fl\quad{}+A_2(1-z)^{\gamma -\alpha-\beta }{}_2F_1(\gamma -\alpha,\gamma -\beta ;
\gamma-\alpha -\beta +1;1-z),\quad
|\arg(1-z)|<\pi,
\end{eqnarray}
where
\begin{equation}
A_1=\frac{\Gamma(\gamma )\Gamma(\gamma -\alpha
-\beta)}{\Gamma(\gamma -\alpha )\Gamma(\gamma -\beta)},\qquad
A_2=\frac{\Gamma(\gamma )\Gamma(\alpha +\beta -\gamma
)}{\Gamma(\alpha)\Gamma(\beta )},
\end{equation}
we get
\begin{eqnarray}
&&\fl{}_2F_1[\hbox{$\scriptstyle{1\over 2}$}(m+
\hbox{$\scriptstyle{1\over 2}$}+{\rm i}\lambda),
\hbox{$\scriptstyle{1\over 2}$}(-m+\hbox{$\scriptstyle{1\over 2}$}+{\rm i}\lambda)
;\hbox{$\scriptstyle{1\over 2}$};{\rm tgh}^2\tau]\nonumber\\
&&\fl\quad{}={\hat A}_1\,{}_2F_1\left[\hbox{$\scriptstyle{1\over 2}$}(m+
\hbox{$\scriptstyle{1\over 2}$}+{\rm i}\lambda),
\hbox{$\scriptstyle{1\over 2}$}(-m+\hbox{$\scriptstyle{1\over 2}$}+{\rm i}\lambda)
;1+{\rm i}\lambda;\frac{1}{\cosh^2\tau}\right]\nonumber\\
&&\fl\qquad{}+{\hat A}_2\cosh^{2{\rm i}\lambda}\tau{}_2F_1\left[
\hbox{$\scriptstyle{1\over 2}$}(-m+
\hbox{$\scriptstyle{1\over 2}$}-{\rm i}\lambda),
\hbox{$\scriptstyle{1\over 2}$}(m+\hbox{$\scriptstyle{1\over 2}$}-{\rm i}\lambda)
;1-{\rm i}\lambda;\frac{1}{\cosh^2\tau}\right],
\end{eqnarray}
where
\numparts
\begin{eqnarray}
{\hat A}_1 &=& \frac{\sqrt{\pi}\Gamma(-{\rm i}\lambda)}{\Gamma[\hbox{$\scriptstyle{1\over 2}$}
(-m+\hbox{$\scriptstyle{1\over 2}$}-{\rm
i}\lambda)]\Gamma[\hbox{$\scriptstyle{1\over
2}$}(m+\hbox{$\scriptstyle{1\over 2}$}-{\rm i}\lambda )]},\\
{\hat A}_2 &=&{\hat A}_1^*= \frac{\sqrt{\pi}\Gamma({\rm i}\lambda)}{\Gamma[
\hbox{$\scriptstyle{1\over 2}$}
(m+\hbox{$\scriptstyle{1\over 2}$}+{\rm
i}\lambda)]\Gamma[\hbox{$\scriptstyle{1\over
2}$}(-m+\hbox{$\scriptstyle{1\over 2}$}+{\rm i}\lambda )]}.
\end{eqnarray}
\endnumparts
Hence, taking into account the relations
\begin{equation}
\lim_{\tau\to\pm\infty}\cosh\tau =\frac{e^{\pm\tau}}{2},\qquad
{}_2F_1(\alpha,\beta ;\gamma;0)=1,
\end{equation}
we obtain the asymptotic form of ${\hat {\cal Y}}^m_\lambda(\tau,\varphi)$.  
Namely, we have
\begin{equation}
\lim_{\tau\to\infty}{\hat{\cal Y}}^m_\lambda(\tau,\varphi)=e^{{\rm
i}m\varphi}{\hat c}_{m\lambda} \cosh^{-\frac{1}{2}}\tau(Be^{-{\rm
i}\lambda\tau}+B^*e^{{\rm i}\lambda\tau}),
\end{equation}
where $|B|=|{\hat A}_1|$.  On demanding that asymptotic functions
are normalized as (see (3.5))
\begin{equation}
\int_{-\infty}^{\infty}\cosh\tau d\tau\int_{0}^{2\pi}d\varphi 
[{\hat {\cal Y}}^m_\lambda(\tau,\varphi)]^*
{\hat {\cal Y}}^{m'}_{\lambda'}(\tau,\varphi)=\delta_{mm'}
\delta(\lambda-\lambda')
\end{equation}
we arrive at the following formula on the normalization constant
\begin{equation}
|{\hat c}_{m\lambda}|^2=\frac{1}{8\pi^2|{\hat A}_1|^2}=\frac{1}{8\pi^3}
\frac{|\Gamma[\hbox{$\scriptstyle{1\over 2}$}
(m+\hbox{$\scriptstyle{1\over 2}$}+{\rm i}\lambda)]\Gamma[\hbox{$\scriptstyle{1\over
2}$}(-m+\hbox{$\scriptstyle{1\over 2}$}+{\rm i}\lambda
)]|^2}{|\Gamma({\rm i}\lambda)|^2}.
\end{equation}

We now normalize the function ${\bar {\cal
Y}}^m_\lambda(\tau,\varphi)$.  Using (5.3) and (5.4) we find
\begin{eqnarray}
&&\fl{\rm tgh}\tau{}_2F_1[\hbox{$\scriptstyle{1\over 2}$}(m+
\hbox{$\scriptstyle{3\over 2}$}+{\rm i}\lambda),
\hbox{$\scriptstyle{1\over 2}$}(-m+\hbox{$\scriptstyle{3\over 2}$}+{\rm i}\lambda)
;\hbox{$\scriptstyle{3\over 2}$};{\rm tgh}^2\tau]\nonumber\\
&&\fl\quad{}={\rm tgh}\tau\left\{{\bar A}_1\,{}_2F_1\left[\hbox{$\scriptstyle{1\over 2}$}(m+
\hbox{$\scriptstyle{3\over 2}$}+{\rm i}\lambda),
\hbox{$\scriptstyle{1\over 2}$}(-m+\hbox{$\scriptstyle{3\over 2}$}+{\rm i}\lambda)
;1+{\rm i}\lambda;\frac{1}{\cosh^2\tau}\right]\right.\nonumber\\
&&\fl\left.\qquad{}+{\bar A}_2\cosh^{2{\rm i}\lambda}\tau{}_2F_1\left[
\hbox{$\scriptstyle{1\over 2}$}(-m+\hbox{$\scriptstyle{3\over 2}$}-{\rm i}\lambda),
\hbox{$\scriptstyle{1\over 2}$}(m+\hbox{$\scriptstyle{3\over 2}$}-{\rm i}\lambda)
;1+{\rm i}\lambda;\frac{1}{\cosh^2\tau}\right]\right\},
\end{eqnarray}
where
\numparts
\begin{eqnarray}
{\bar A}_1 &=& \frac{\sqrt{\pi}}{2}\frac{\Gamma(-{\rm i}\lambda)}{\Gamma[
\hbox{$\scriptstyle{1\over 2}$}
(-m+\hbox{$\scriptstyle{3\over 2}$}-{\rm
i}\lambda)]\Gamma[\hbox{$\scriptstyle{1\over
2}$}(m+\hbox{$\scriptstyle{3\over 2}$}-{\rm i}\lambda )]},\\
{\bar A}_2 &=&({\bar A}_1)^*= \frac{\sqrt{\pi}}{2}\frac{\Gamma({\rm i}\lambda)}{\Gamma[
\hbox{$\scriptstyle{1\over 2}$}
(m+\hbox{$\scriptstyle{3\over 2}$}+{\rm
i}\lambda)]\Gamma[\hbox{$\scriptstyle{1\over
2}$}(-m+\hbox{$\scriptstyle{3\over 2}$}+{\rm i}\lambda )]}.
\end{eqnarray}
\endnumparts
Proceeding analogously as in the case of the normalization of the
function ${\hat{\cal Y}}^m_\lambda(\tau,\varphi)$, and making use of
of the limit $\lim_{\tau\to\pm\infty}{\rm tgh}\tau=\pm1$, we obtain
\begin{equation}
|{\bar c}_{m\lambda}|^2=\frac{1}{8\pi^2|{\bar A}_1|^2}=\frac{1}{2\pi^3}
\frac{|\Gamma[\hbox{$\scriptstyle{1\over 2}$}
(m+\hbox{$\scriptstyle{3\over 2}$}+{\rm i}\lambda)]\Gamma[\hbox{$\scriptstyle{1\over
2}$}(-m+\hbox{$\scriptstyle{3\over 2}$}+{\rm i}\lambda
)]|^2}{|\Gamma({\rm i}\lambda)|^2}.
\end{equation}

Taking into account (5.1), (5.2), (5.10), and (5.13) we can finally
write down the desired normalized pseudospherical functions
\begin{eqnarray}
\fl{\hat{\cal Y}}^m_\lambda(\tau,\varphi)={\hat c}_m \frac{e^{{\rm
i}m\varphi}}{2\sqrt{2}\pi^{\frac{3}{2}}}
\frac{|\Gamma[\hbox{$\scriptstyle{1\over 2}$}
(m+\hbox{$\scriptstyle{1\over 2}$}+{\rm i}\lambda)]\Gamma[\hbox{$\scriptstyle{1\over
2}$}(-m+\hbox{$\scriptstyle{1\over 2}$}+{\rm i}\lambda
)]|}{|\Gamma({\rm i}\lambda)|}
\cosh^{-(\frac{1}{2}+{\rm i}\lambda)}\tau\nonumber\\
\fl\qquad\qquad\quad\times\,{}_2F_1[\hbox{$\scriptstyle{1\over 2}$}(m+
\hbox{$\scriptstyle{1\over 2}$}+{\rm i}\lambda),
\hbox{$\scriptstyle{1\over 2}$}(-m+\hbox{$\scriptstyle{1\over 2}$}+{\rm i}\lambda)
;\hbox{$\scriptstyle{1\over 2}$};{\rm tgh}^2\tau],\\
\fl{\bar{\cal Y}}^m_\lambda(\tau,\varphi)={\bar c}_m \frac{e^{{\rm
i}m\varphi}}{\sqrt{2}\pi^{\frac{3}{2}}}
\frac{|\Gamma[\hbox{$\scriptstyle{1\over 2}$}
(m+\hbox{$\scriptstyle{3\over 2}$}+{\rm i}\lambda)]\Gamma[\hbox{$\scriptstyle{1\over
2}$}(-m+\hbox{$\scriptstyle{3\over 2}$}+{\rm i}\lambda
)]|}{|\Gamma({\rm i}\lambda)|}
\cosh^{-(\frac{1}{2}+{\rm i}\lambda)}\tau{\rm tgh}\tau\nonumber\\
\fl\qquad\qquad\quad\times\,{}_2F_1[\hbox{$\scriptstyle{1\over 2}$}
(m+\hbox{$\scriptstyle{3\over 2}$}+{\rm i}\lambda),
\hbox{$\scriptstyle{1\over 2}$}(-m+\hbox{$\scriptstyle{3\over 2}$}+{\rm i}\lambda)
;\hbox{$\scriptstyle{3\over 2}$};{\rm tgh}^2\tau].
\end{eqnarray}
where ${\hat c }_m$ and ${\bar c}_m$ are phases that will be fixed
in the next section.  We point out that obtained normalization of
pseudospherical functions provides the most probably new nontrivial
identities for the hypergeometric functions.
\subsection{The Fock space structure of the continuous principal series
representation}
Although in the case of the continuous series representation there
is neither lowest nor highest weight states, nevertheless there is
still possibility, in view of (2.6), of generation of states with
arbitrary $m$ from some fixed one by means of the ladder operators
$K_\pm$. This possibility is the subject of this section.  Now, it
seems that the most natural candidates for such a distinguished
``vacuum vector'' in the case with integer $m$, is the state with
$m=0$.  Taking into account (2.6) we find for $k=-\frac{1}{2}+{\rm
i}\lambda$ 
\numparts
\begin{eqnarray}
K_+|\lambda,m\rangle &=& \sqrt{(m+\hbox{$\scriptstyle{1\over
2}$})^2+\lambda^2}|\lambda,m+1\rangle,\\
K_-|\lambda,m\rangle &=& \sqrt{(m-\hbox{$\scriptstyle{1\over
2}$})^2+\lambda^2}|\lambda,m-1\rangle,
\end{eqnarray}
\endnumparts
where $|\lambda,m\rangle\equiv|-\hbox{$\scriptstyle{1\over 2}$}+{\rm
i}\lambda,m \rangle$.  An easy calculation based on (5.16a) shows that
\begin{equation}
|\lambda,m\rangle =
\frac{1}{\prod_{r=1}^m\sqrt{(\frac{2r-1}{2})^2+\lambda^2}}K_+^m
|\lambda,0\rangle,\qquad m=1,\,2,\,\ldots.
\end{equation}
We remark that the fraction from (5.17) can be expressed by the
gamma functions.  Namely, using the identity \cite{21}
\begin{equation}
\frac{\Gamma(m+\hbox{$\scriptstyle{1\over 2}$}-{\rm i}\lambda)}
{\Gamma(-m+\hbox{$\scriptstyle{1\over 2}$}-{\rm i}\lambda)} =
(-1)^m\prod_{r=1}^m\left[\left(\frac{2r-1}{2}\right)^2+\lambda^2\right]
\end{equation}
and \cite{17}
\begin{equation}
\Gamma(z)=\frac{2^{z-1}}{\sqrt{\pi}}\Gamma(\hbox{$\scriptstyle{1\over
2}$}z)\Gamma[\hbox{$\scriptstyle{1\over 2}$}(z+1)],
\end{equation}
one can show that
\numparts
\begin{eqnarray}
\frac{1}{\prod_{r=1}^m\sqrt{(\frac{2r-1}{2})^2+\lambda^2}} &=&
2^{-m}\frac{|\Gamma[\hbox{$\scriptstyle{1\over 2}$}(-m+
\hbox{$\scriptstyle{1\over 2}$}+{\rm i}\lambda)]|}
{|\Gamma[\hbox{$\scriptstyle{1\over
2}$}(m+\hbox{$\scriptstyle{1\over 2}$}+{\rm i}\lambda)]|},\\
&=& 2^{-m}\frac{|\Gamma[\hbox{$\scriptstyle{1\over 2}$}(-m+
\hbox{$\scriptstyle{3\over 2}$}+{\rm i}\lambda)]|}
{|\Gamma[\hbox{$\scriptstyle{1\over
2}$}(m+\hbox{$\scriptstyle{3\over 2}$}+{\rm i}\lambda)]|}.
\end{eqnarray}
\endnumparts
We remark that the right-hand side of (5.18) is usually designated
by $Z^m_\lambda$.  Now, we have in our disposal two possibilities to
choose the function representing vector $|\lambda,0\rangle$, namely
${\hat {\cal Y}}^0_\lambda$, and ${\bar {\cal Y}}^0_\lambda$ (see
(5.1) and (5.2)), and therefore two different sequences of 
pseudospherical functions generated via (5.17).  Let us designate 
these sequences by ${{\cal Y}_1}^m_\lambda$ and ${{\cal Y}_2}^m_\lambda$, 
respectively.  In order to identify elements of ${{\cal Y}_1}^m_\lambda$ and 
${{\cal Y}_2}^m_\lambda$ we first point out that the functions
${\hat {\cal Y}}^m_\lambda$ are even and the functions ${\bar {\cal Y}}^m_\lambda$
are odd functions of $\tau$ for arbitrary $m$.  Furthermore, it
follows immediately from the form (2.17a) of the operator $K_+$ that
it maps an even function to odd one and vice versa. Therefore, the
sequence ${{\cal Y}_1}^m_\lambda$ is of the form
\begin{equation}
{\hat {\cal Y}}^0_\lambda,\,{\bar {\cal Y}}^1_\lambda,\,{\hat {\cal
Y}}^2_\lambda,\,\ldots ,\,{\hat {\cal Y}}^{2n}_\lambda,\,{\bar {\cal
Y}}^{2n+1}_\lambda,\,\ldots ,
\end{equation}
that is
\begin{equation}
{{\cal Y}_1}^m_\lambda = \cases{{\hat {\cal Y}}^m_\lambda, &for $m=2n$,\cr
{\bar {\cal Y}}^m_\lambda, &for $m=2n+1,\qquad n=0,\,1,\,\ldots$}
\end{equation}
Analogously, the sequence ${{\cal Y}_2}^m_\lambda$ is given by
\begin{equation}
{\bar {\cal Y}}^0_\lambda,\,{\hat {\cal Y}}^1_\lambda,\,{\bar {\cal
Y}}^2_\lambda,\,\ldots ,\,{\bar {\cal Y}}^{2n}_\lambda,\,{\hat {\cal
Y}}^{2n+1}_\lambda,\,\ldots ,
\end{equation}
i.e.
\begin{equation}
{{\cal Y}_2}^m_\lambda = \cases{{\bar {\cal Y}}^m_\lambda, &for $m=2n$,\cr
{\hat {\cal Y}}^m_\lambda, &for $m=2n+1,\qquad n=0,\,1,\,\ldots$}
\end{equation}
Surprisingly, this structure of the principal series was not, to our
best knowledge, discussed in the literature.  We stress that in the
case of the discrete series all functions, even and odd ones, are
elements of the only one sequence of pseudospherical functions
starting with the unique highest or lowest weight state (for the
series $D_+$ see (3.16)).  We now derive the counterpart of the
formula (3.15) in the case of the continuous series.  We begin with
the case of the functions ${{\cal Y}_1}^m_\lambda$.  To this end we
first write down the function ${\hat {\cal
Y}}^0_\lambda(\tau,\varphi)$ following directly from (5.14)
\begin{equation}
\fl{\hat{\cal Y}}_\lambda^0(\tau,\varphi)=\frac{1}{2\sqrt{2}\pi^{\frac{3}{2}}}
\frac{|\Gamma[\hbox{$\scriptstyle{1\over 2}$}
(\hbox{$\scriptstyle{1\over 2}$}+{\rm i}\lambda)]|^2}
{|\Gamma({\rm i}\lambda)|}\cosh^{-(\frac{1}{2}+{\rm i}\lambda)}\tau
{}_2F_1[\hbox{$\scriptstyle{1\over 2}$}(\hbox{$\scriptstyle{1\over 2}$}
+{\rm i}\lambda),\hbox{$\scriptstyle{1\over 2}$}(\hbox{$\scriptstyle{1\over 2}$}
+{\rm i}\lambda);\hbox{$\scriptstyle{1\over 2}$};{\rm tgh}^2\tau],
\end{equation}
where we choose, without loose of generality, the phase so ${\hat
c_0}=1$.  Using the identity \cite{21}
\begin{equation}
{}_2F_1(\alpha,\beta ;\gamma;z) =
(1-z)^{-\alpha}{}_2F_1\left(\alpha,\gamma -\beta ;
\gamma;\frac{z}{z-1}\right)
\end{equation}
we obtain the following equivalent form of the function ${\hat{\cal
Y}}^0_\lambda$ such that
\begin{equation}
\fl {\hat{\cal Y}}_\lambda^0(\tau,\varphi)=\frac{1}{2\sqrt{2}\pi^{\frac{3}{2}}}
\frac{|\Gamma[\hbox{$\scriptstyle{1\over 2}$}
(\hbox{$\scriptstyle{1\over 2}$}+{\rm i}\lambda)]|^2}
{|\Gamma({\rm i}\lambda)|}
{}_2F_1[\hbox{$\scriptstyle{1\over 2}$}(\hbox{$\scriptstyle{1\over 2}$}
+{\rm i}\lambda),\hbox{$\scriptstyle{1\over 2}$}(\hbox{$\scriptstyle{1\over 2}$}
-{\rm i}\lambda);\hbox{$\scriptstyle{1\over 2}$};-\sinh^2\tau].
\end{equation}
It follows immediately from (3.8) and (5.17) that the
pseudospherical functions ${{\cal Y}_1}^m_\lambda$ are given by
\begin{eqnarray}
\fl{{\cal Y}_1}^m_\lambda(\tau,\varphi) = (-1)^m\frac{1}{2\sqrt{2}\pi^{\frac{3}{2}}}
\frac{|\Gamma[\hbox{$\scriptstyle{1\over 2}$}
(\hbox{$\scriptstyle{1\over 2}$}+{\rm i}\lambda)]|^2}
{|\Gamma({\rm i}\lambda)|}\frac{1}{\prod_{r=1}^m\sqrt{(\frac{2r-1}{2})^2+\lambda^2}}
e^{{\rm i}m\varphi}\cosh^m\tau\nonumber\\
\qquad\qquad\quad\fl\times\,\frac{d^m}{d(\sinh\tau)^m}{}_2F_1[\hbox{$\scriptstyle{1\over 2}$}
(\hbox{$\scriptstyle{1\over 2}$}
+{\rm i}\lambda),\hbox{$\scriptstyle{1\over 2}$}(\hbox{$\scriptstyle{1\over 2}$}
-{\rm i}\lambda);\hbox{$\scriptstyle{1\over 2}$};-\sinh^2\tau].
\end{eqnarray}
Notice that the relation (5.28) can be written as (compare (3.10))
\begin{equation}
\fl{{\cal Y}_1}^m_\lambda(\tau,\varphi) = 
\frac{1}{2\sqrt{2}\pi^{\frac{3}{2}}}\frac{1}{\prod_{r=1}^m\sqrt{(\frac{2r-1}{2})^2+\lambda^2}}
e^{{\rm i}m\varphi}{{\cal P}_1}^m_\lambda(\sinh\tau),
\end{equation}
where ${{\cal P}_1}^m_\lambda(x)$ are the plausible pseudospherical
counterparts of the Legendre functions defined as
\begin{equation}
\fl{{\cal P}_1}^m_\lambda(x)=(-1)^m\frac{|\Gamma[\hbox{$\scriptstyle{1\over 2}$}
(\hbox{$\scriptstyle{1\over 2}$}+{\rm i}\lambda)]|^2}
{|\Gamma({\rm i}\lambda)|}(1+x^2)^{\frac{m}{2}}\frac{d^m}{dx^m}
{}_2F_1[\hbox{$\scriptstyle{1\over 2}$}
(\hbox{$\scriptstyle{1\over 2}$}
+{\rm i}\lambda),\hbox{$\scriptstyle{1\over 2}$}(\hbox{$\scriptstyle{1\over 2}$}
-{\rm i}\lambda);\hbox{$\scriptstyle{1\over 2}$};-x^2].
\end{equation}
Now, taking into account (3.12) we get
\begin{eqnarray}
\fl\frac{d^m}{dx^m}{}_2F_1[\hbox{$\scriptstyle{1\over 2}$}
(\hbox{$\scriptstyle{1\over 2}$}
+{\rm i}\lambda),\hbox{$\scriptstyle{1\over 2}$}(\hbox{$\scriptstyle{1\over 2}$}
-{\rm i}\lambda);\hbox{$\scriptstyle{1\over 2}$};-x^2]\nonumber\\
\fl\qquad{}=m!\sum_{r=0}^{[\frac{m}{2}]}\frac{(2x)^{m-2r}}{r!(m-2r)!}\left(
\frac{d^{m-r}}{dx^{m-r}}{}_2F_1[\hbox{$\scriptstyle{1\over 2}$}
(\hbox{$\scriptstyle{1\over 2}$}
+{\rm i}\lambda),\hbox{$\scriptstyle{1\over 2}$}(\hbox{$\scriptstyle{1\over 2}$}
-{\rm i}\lambda);\hbox{$\scriptstyle{1\over 2}$};-x]\right)\Big\vert_{x=x^2}.
\end{eqnarray}
Hence, using the identity \cite{20}
\begin{equation}
\fl\frac{d^n}{dz^n}{}_2F_1(\alpha,\beta ;\gamma;z) =
\frac{\Gamma(\alpha
+n)\Gamma(\beta+n)\Gamma(\gamma)}{\Gamma(\alpha)\Gamma(\beta)\Gamma(\gamma
+n)}{}_2F_1(\alpha+n,\beta+n;\gamma+n;z)
\end{equation}
we find
\begin{eqnarray}
\fl{{\cal P}_1}^m_\lambda(x) = \frac{\sqrt{\pi}m!}{|\Gamma({\rm
i}\lambda)|}(1+x^2)^{\frac{m}{2}}\sum_{r=0}^{[\frac{m}{2}]}(-1)^r\frac{(2x)^{m-2r}}{r!(m-2r)!}
\frac{|\Gamma[\hbox{$\scriptstyle{1\over
2}$}(\hbox{$\scriptstyle{1\over 2}$}-{\rm
i}\lambda)+m-r]|^2}{\Gamma(\hbox{$\scriptstyle{1\over
2}$}+m-r)}\nonumber\\
\fl\qquad\qquad\times\,{}_2F_1[\hbox{$\scriptstyle{1\over 2}$}
(\hbox{$\scriptstyle{1\over 2}$}
+{\rm i}\lambda)+m-r,\hbox{$\scriptstyle{1\over 2}$}(\hbox{$\scriptstyle{1\over 2}$}
-{\rm i}\lambda)+m-r;\hbox{$\scriptstyle{1\over 2}$}+m-r;-x^2].
\end{eqnarray}
Substituting $x=\sinh\tau$ and making use of (5.26) we obtain
\begin{eqnarray}
\fl{{\cal P}_1}^m_\lambda(\sinh\tau) = \frac{\sqrt{\pi}m!}{|\Gamma({\rm
i}\lambda)|}\cosh^{-(\frac{1}{2}+{\rm i}\lambda)}\tau\sum_{r=0}^{[\frac{m}{2}]}(-1)^r\frac{(2{\rm
tgh}\tau)^{m-2r}}{r!(m-2r)!}
\frac{|\Gamma[\hbox{$\scriptstyle{1\over
2}$}(\hbox{$\scriptstyle{1\over 2}$}-{\rm
i}\lambda)+m-r]|^2}{\Gamma(\hbox{$\scriptstyle{1\over
2}$}+m-r)}\nonumber\\
\fl\qquad\qquad\qquad\times\,{}_2F_1[\hbox{$\scriptstyle{1\over 2}$}
(\hbox{$\scriptstyle{1\over 2}$}
+{\rm i}\lambda)+m-r,\hbox{$\scriptstyle{1\over 2}$}(\hbox{$\scriptstyle{1\over 2}$}
+{\rm i}\lambda);\hbox{$\scriptstyle{1\over 2}$}+m-r;{\rm tgh}^2\tau].
\end{eqnarray}
Therefore, the pseudospherical functions ${{\cal
Y}_1}^m_\lambda(\tau,\varphi)$
can be written in the form
\begin{eqnarray}
\fl{{\cal Y}_1}^m_\lambda(\tau,\varphi) = \frac{m!}{2\sqrt{2}\pi|\Gamma({\rm
i}\lambda)|}\frac{1}{\prod_{r=1}^m\sqrt{(\frac{2r-1}{2})^2+\lambda^2}}
e^{{\rm i}m\varphi}\cosh^{-(\frac{1}{2}+{\rm
i}\lambda)}\tau\sum_{r=0}^{[\frac{m}{2}]}(-1)^r\frac{(2{\rm
tgh}\tau)^{m-2r}}{r!(m-2r)!}\nonumber\\
\fl\qquad\times\,\frac{|\Gamma[\hbox{$\scriptstyle{1\over
2}$}(\hbox{$\scriptstyle{1\over 2}$}-{\rm
i}\lambda)+m-r]|^2}{\Gamma(\hbox{$\scriptstyle{1\over
2}$}+m-r)}{}_2F_1[\hbox{$\scriptstyle{1\over 2}$}
(\hbox{$\scriptstyle{1\over 2}$}
+{\rm i}\lambda)+m-r,\hbox{$\scriptstyle{1\over 2}$}(\hbox{$\scriptstyle{1\over 2}$}
+{\rm i}\lambda);\hbox{$\scriptstyle{1\over 2}$}+m-r;{\rm tgh}^2\tau],
\end{eqnarray}
where $m=1,\,2,\,\ldots$.

We now discuss the pseudospherical functions ${{\cal
Y}_2}^m_\lambda(\tau,\varphi)$.  From (5.15) we immediately find
that the odd ``vacuum vector'' ${\bar {\cal Y}}^0_\lambda(\tau,\varphi)$ is
\begin{equation}
\fl{\bar{\cal Y}}_\lambda^0(\tau,\varphi)=\frac{1}{\sqrt{2}\pi^{\frac{3}{2}}}
\frac{|\Gamma[\hbox{$\scriptstyle{1\over 2}$}
(\hbox{$\scriptstyle{3\over 2}$}+{\rm i}\lambda)]|^2}
{|\Gamma({\rm i}\lambda)|}\cosh^{-(\frac{1}{2}+{\rm
i}\lambda)}\tau{\rm tgh}\tau
{}_2F_1[\hbox{$\scriptstyle{1\over 2}$}(\hbox{$\scriptstyle{3\over 2}$}
+{\rm i}\lambda),\hbox{$\scriptstyle{1\over 2}$}(\hbox{$\scriptstyle{3\over 2}$}
+{\rm i}\lambda);\hbox{$\scriptstyle{3\over 2}$};{\rm tgh}^2\tau],
\end{equation}
where we set ${\bar c}=1$.  Using (5.26) we get
\begin{equation}
\fl {\bar{\cal Y}}_\lambda^0(\tau,\varphi)=\frac{1}{\sqrt{2}\pi^{\frac{3}{2}}}
\frac{|\Gamma[\hbox{$\scriptstyle{1\over 2}$}
(\hbox{$\scriptstyle{3\over 2}$}+{\rm i}\lambda)]|^2}
{|\Gamma({\rm i}\lambda)|}\sinh\tau
{}_2F_1[\hbox{$\scriptstyle{1\over 2}$}(\hbox{$\scriptstyle{3\over 2}$}
+{\rm i}\lambda),\hbox{$\scriptstyle{1\over 2}$}(\hbox{$\scriptstyle{3\over 2}$}
-{\rm i}\lambda);\hbox{$\scriptstyle{3\over 2}$};-\sinh^2\tau].
\end{equation}
Taking into account (5.17) we arrive at the following formula on the
pseudospherical functions ${{\cal Y}_2}^m_\lambda(\tau,\varphi)$:
\begin{eqnarray}
\fl{{\cal Y}_2}^m_\lambda(\tau,\varphi) = (-1)^m\frac{1}{\sqrt{2}\pi^{\frac{3}{2}}}
\frac{|\Gamma[\hbox{$\scriptstyle{1\over 2}$}
(\hbox{$\scriptstyle{3\over 2}$}+{\rm i}\lambda)]|^2}
{|\Gamma({\rm i}\lambda)|}\frac{1}{\prod_{r=1}^m\sqrt{(\frac{2r-1}{2})^2+\lambda^2}}
e^{{\rm i}m\varphi}\cosh^m\tau\nonumber\\
\qquad\qquad\quad\fl\times\,\frac{d^m}{d(\sinh\tau)^m}\sinh\tau
{}_2F_1[\hbox{$\scriptstyle{1\over 2}$}
(\hbox{$\scriptstyle{3\over 2}$}
+{\rm i}\lambda),\hbox{$\scriptstyle{1\over 2}$}(\hbox{$\scriptstyle{3\over 2}$}
-{\rm i}\lambda);\hbox{$\scriptstyle{3\over 2}$};-\sinh^2\tau].
\end{eqnarray}
Introducing the counterparts of the Legendre functions ${{\cal P}_2}^m_\lambda(x)$
we can write (5.38) in the form
\begin{equation}
\fl{{\cal Y}_2}^m_\lambda(\tau,\varphi) = 
\frac{1}{\sqrt{2}\pi^{\frac{3}{2}}}\frac{1}{\prod_{r=1}^m\sqrt{(\frac{2r-1}{2})^2+\lambda^2}}
e^{{\rm i}m\varphi}{{\cal P}_2}^m_\lambda(\sinh\tau),
\end{equation}
where
\begin{equation}
\fl{{\cal P}_2}^m_\lambda(x)=(-1)^m\frac{|\Gamma[\hbox{$\scriptstyle{1\over 2}$}
(\hbox{$\scriptstyle{3\over 2}$}+{\rm i}\lambda)]|^2}
{|\Gamma({\rm i}\lambda)|}(1+x^2)^{\frac{m}{2}}\frac{d^m}{dx^m}
x{}_2F_1[\hbox{$\scriptstyle{1\over 2}$}
(\hbox{$\scriptstyle{3\over 2}$}
+{\rm i}\lambda),\hbox{$\scriptstyle{1\over 2}$}(\hbox{$\scriptstyle{3\over 2}$}
-{\rm i}\lambda);\hbox{$\scriptstyle{3\over 2}$};-x^2].
\end{equation}
Eqs.\ (5.40), (3.12) and (5.32) taken together yield
\begin{eqnarray}
\fl{{\cal P}_2}^m_\lambda(x) = \frac{\sqrt{\pi}(m+1)!}{4|\Gamma({\rm
i}\lambda)|}(1+x^2)^{\frac{m}{2}}\sum_{r=0}^{[\frac{m+1}{2}]}(-1)^r\frac{(2x)^{m+1-2r}}{r!(m+1-2r)!}
\frac{|\Gamma[\hbox{$\scriptstyle{1\over
2}$}(\hbox{$\scriptstyle{3\over 2}$}+{\rm
i}\lambda)+m-r]|^2}{\Gamma(\hbox{$\scriptstyle{3\over
2}$}+m-r)}\nonumber\\
\fl\qquad\qquad\times\,{}_2F_1[\hbox{$\scriptstyle{1\over 2}$}
(\hbox{$\scriptstyle{3\over 2}$}
+{\rm i}\lambda)+m-r,\hbox{$\scriptstyle{1\over
2}$}(\hbox{$\scriptstyle{3\over 2}$}
-{\rm i}\lambda)+m-r;\hbox{$\scriptstyle{3\over 2}$}+m-r;-x^2].
\end{eqnarray}
Hence, using (5.26) we get
\begin{eqnarray}
\fl{{\cal P}_2}^m_\lambda(\sinh\tau) = \frac{\sqrt{\pi}(m+1)!}{4|\Gamma({\rm
i}\lambda)|}\cosh^{-(\frac{1}{2}+{\rm
i}\lambda)}\tau\sum_{r=0}^{[\frac{m+1}{2}]}(-1)^r\frac{(2{\rm
tgh}\tau)^{m+1-2r}}{r!(m+1-2r)!}
\frac{|\Gamma[\hbox{$\scriptstyle{1\over
2}$}(\hbox{$\scriptstyle{3\over 2}$}+{\rm
i}\lambda)+m-r]|^2}{\Gamma(\hbox{$\scriptstyle{3\over
2}$}+m-r)}\nonumber\\
\fl\qquad\qquad\qquad\times\,{}_2F_1[\hbox{$\scriptstyle{1\over 2}$}
(\hbox{$\scriptstyle{3\over 2}$}
+{\rm i}\lambda)+m-r,\hbox{$\scriptstyle{1\over
2}$}(\hbox{$\scriptstyle{3\over 2}$}
+{\rm i}\lambda);\hbox{$\scriptstyle{3\over 2}$}+m-r;{\rm tgh}^2\tau].
\end{eqnarray}
An immediate consequence of (5.39) and (5.42) is the following
formula on the pseudospherical functions ${{\cal
Y}_2}^m_\lambda(\tau,\varphi)$:
\begin{eqnarray}
\fl{{\cal Y}_2}^m_\lambda(\tau,\varphi) = \frac{(m+1)!}{4\sqrt{2}\pi|\Gamma({\rm
i}\lambda)|}\frac{1}{\prod_{r=1}^m\sqrt{(\frac{2r-1}{2})^2+\lambda^2}}
e^{{\rm i}m\varphi}\cosh^{-(\frac{1}{2}+{\rm
i}\lambda)}\tau\sum_{r=0}^{[\frac{m+1}{2}]}(-1)^r\frac{(2{\rm
tgh}\tau)^{m+1-2r}}{r!(m+1-2r)!}\nonumber\\
\fl\qquad\times\,\frac{|\Gamma[\hbox{$\scriptstyle{1\over
2}$}(\hbox{$\scriptstyle{3\over 2}$}+{\rm
i}\lambda)+m-r]|^2}{\Gamma(\hbox{$\scriptstyle{3\over
2}$}+m-r)}{}_2F_1[\hbox{$\scriptstyle{1\over 2}$}
(\hbox{$\scriptstyle{3\over 2}$}
+{\rm i}\lambda)+m-r,\hbox{$\scriptstyle{1\over
2}$}(\hbox{$\scriptstyle{3\over 2}$}
+{\rm i}\lambda);\hbox{$\scriptstyle{3\over 2}$}+m-r;{\rm tgh}^2\tau],
\end{eqnarray}
where $m=1,\,2,\,\ldots$.  An advantage of the relations (5.35) and
(5.43) is that they enable to fix
with the help of the Gauss relations between contiguous functions
\cite{21} the phases of the even and odd pseudospherical functions
(5.14) and (5.15).  Namely, we find
\begin{equation}
{\hat c}_m = \cases{(-1)^\frac{m}{2}, &for $m=2n$,\cr
(-1)^\frac{m+1}{2}, &for $m=2n+1,\qquad n=0,\,1,\,\ldots$}
\end{equation}
and
\begin{equation}
{\bar c}_m = \cases{(-1)^\frac{m}{2}, &for $m=2n$,\cr
(-1)^\frac{m-1}{2}, &for $m=2n+1,\qquad n=0,\,1,\,\ldots$}
\end{equation}
We stress that the phases cannot be fixed without knowledge of the
infinite sequence of pseudospherical functions generated by means of
the ladder operators.

We finally point out that an immediate consequence of the relation 
\begin{equation}
|\lambda,-m\rangle =
\frac{1}{\prod_{r=1}^m\sqrt{(\frac{2r-1}{2})^2+\lambda^2}}K_-^m
|\lambda,0\rangle,\qquad m=1,\,2,\,\ldots.
\end{equation}
following directly from (5.16b) is the formula on ${{\cal Y}_1}^{-m}_\lambda$
and ${{\cal Y}_2}^{-m}_\lambda$ such that
\begin{equation}
{{\cal Y}_{1,2}}^{-m}_\lambda(\tau,\varphi) = (-1)^m[{{\cal
Y}_{1,2}}^m_\lambda(\tau,\varphi)]^*,\qquad m=1,\,2,\,\ldots .
\end{equation}
\subsection{Pseudospherical functions with half-integer {\em m}}
In this section we study the pseudospherical functions ${{\cal Y}_1}^m_\lambda$
and ${{\cal Y}_2}^m_\lambda$ in the case of the continuous
principal series and half-integer $m$.  As far as we are aware the
pseudospherical functions with half-integer $m$ were not discussed
in the literature.  We now return to (5.14).  Taking into account
the identity \cite{20}
\begin{equation}
{}_2F_1(\alpha+\hbox{$\scriptstyle{1\over 2}$},\alpha
;\hbox{$\scriptstyle{1\over 2}$};z) =
\hbox{$\scriptstyle{1\over 2}$}(1+\sqrt{z})^{-2\alpha}+
\hbox{$\scriptstyle{1\over 2}$}(1-\sqrt{z})^{-2\alpha}
\end{equation}
as well as the following properties of the gamma function \cite{17}:
\begin{eqnarray}
|\Gamma(\hbox{$\scriptstyle{1\over 2}$}+{\rm i}x)|^2 &=&
\frac{\pi}{\cosh\pi x},\\
|\Gamma({\rm i}x)|^2 &=& \frac{\pi}{x\sinh\pi x},
\end{eqnarray}
and setting in (5.14) $m=\frac{1}{2}$, and ${\hat c}_\frac{1}{2}=1$,
we get the normalized pseudospherical functions
\begin{equation}
{\hat{\cal Y}}^\frac{1}{2}_\lambda(\tau,\varphi) =
\frac{1}{\sqrt{2}\pi}e^{\frac{{\rm i}\varphi}{2}}
\frac{\cos\lambda\tau}{\sqrt{\cosh\tau}}.
\end{equation}
Therefore, in opposition to (5.25) the ``vacuum vector'' for
$m=\frac{1}{2}$ is represented by an elementary function.
Similarly, using (5.48) and the identity \cite{20}
\begin{eqnarray}
&&(-1)^n\frac{\Gamma(\alpha +n)\Gamma(\gamma -\beta+n)\Gamma(\gamma)}
{\Gamma(\alpha)\Gamma(\gamma-\beta)\Gamma(\gamma
+n)}(1-z)^{\alpha -1}{}_2F_1(\alpha+n,\beta;\gamma+n;z)\nonumber\\
&&\qquad{}=\frac{d^n}{dz^n}[(1-z)^{\alpha +n-1}{}_2F_1(\alpha,\beta
;\gamma;z)],
\end{eqnarray}
we obtain from (5.15)
\begin{equation}
{\bar{\cal Y}}^\frac{1}{2}_\lambda(\tau,\varphi) =
\frac{1}{\sqrt{2}\pi}e^{\frac{{\rm i}\varphi}{2}}
\frac{\sin\lambda\tau}{\sqrt{\cosh\tau}}.
\end{equation}
We now study the Fock space structure of the continuous principal
series representation for half-integer $m$.  We begin with the
functions ${{\cal Y}_1}^m_\lambda(\tau,\varphi)$.  From (5.16a) we
obtain for $m=l+\frac{1}{2}$
\begin{equation}
|\lambda,l+\hbox{$\scriptstyle{1\over 2}$}\rangle =
\frac{1}{\prod_{r=1}^l\sqrt{r^2+\lambda^2}}K_+^l
|\lambda,\hbox{$\scriptstyle{1\over 2}$}\rangle,\qquad l=1,\,2,\,\ldots.
\end{equation}
Hence, taking into account (3.8) and (5.51) we find
\begin{equation}
\fl{{\cal Y}_1}^{l+\frac{1}{2}}_\lambda(\tau,\varphi) =
\frac{(-1)^l}{\sqrt{2}\pi}\frac{1}{\prod_{r=1}^l\sqrt{r^2+\lambda^2}}
e^{{\rm
i}(l+\frac{1}{2}\varphi)}\cosh^{l+\frac{1}{2}}\tau\frac{d^l}{d(\sinh\tau)^l}
\frac{\cos\lambda\tau}{\cosh\tau}.
\end{equation}
Finally, using the identity
\begin{equation}
\cos\lambda\tau = \frac{1}{2}[(\sqrt{1+\sinh^2\tau}+\sinh\tau)^{{\rm
i}\lambda}+(\sqrt{1+\sinh^2\tau}-\sinh\tau)^{{\rm
i}\lambda}],
\end{equation}
we obtain the formula 
\begin{eqnarray}
\fl{{\cal Y}_1}^{l+\frac{1}{2}}_\lambda(\tau,\varphi) =
\frac{(-1)^l}{2\sqrt{2}\pi}\frac{1}{\prod_{r=1}^l\sqrt{r^2+\lambda^2}}
e^{{\rm
i}(l+\frac{1}{2})\varphi}\cosh^{l+\frac{1}{2}}\tau\nonumber\\
\fl\times\,\frac{d^l}{d(\sinh\tau)^l}\frac{1}{\sqrt{1+\sinh^2\tau}}
[(\sqrt{1+\sinh^2\tau}+\sinh\tau)^{{\rm
i}\lambda}+(\sqrt{1+\sinh^2\tau}-\sinh\tau)^{{\rm i}\lambda}].
\end{eqnarray}
In view of the form of (5.57) the counterparts ${\cal T}^l_\lambda$
of the Legendre functions such that
\begin{equation}
{{\cal Y}_1}^{l+\frac{1}{2}}_\lambda(\tau,\varphi) = 
\frac{1}{2\sqrt{2}\pi}\frac{1}{\prod_{r=1}^l\sqrt{r^2+\lambda^2}}
e^{{\rm i}(l+\frac{1}{2})\varphi}{\cal T}^l_\lambda(\sinh\tau)
\end{equation}
are defined by
\begin{equation}
\fl{\cal T}^l_\lambda(x) =
(-1)^l(1+x^2)^{\frac{1}{2}(l+\frac{1}{2})}\frac{d^l}{dx^l}\frac{1}{\sqrt{1+x^2}}
[(\sqrt{1+x^2}+x)^{{\rm i}\lambda}+(\sqrt{1+x^2}-x)^{{\rm i}\lambda}].
\end{equation}
We now study the functions ${{\cal Y}_2}^m_\lambda(\tau,\varphi)$ for
half-integer $m$.  Eqs.\ (5.54), (3.8) and (5.53) taken together imply
\begin{equation}
\fl{{\cal Y}_2}^{l+\frac{1}{2}}_\lambda(\tau,\varphi) =
\frac{(-1)^l}{\sqrt{2}\pi}\frac{1}{\prod_{r=1}^l\sqrt{r^2+\lambda^2}}
e^{{\rm
i}(l+\frac{1}{2}\varphi)}\cosh^{l+\frac{1}{2}}\tau\frac{d^l}{d(\sinh\tau)^l}
\frac{\sin\lambda\tau}{\cosh\tau}.
\end{equation}
An immediate consequence of the identity
\begin{equation}
\sin\lambda\tau = \frac{1}{2{\rm i}}[(\sqrt{1+\sinh^2\tau}+\sinh\tau)^{{\rm
i}\lambda}-(\sqrt{1+\sinh^2\tau}-\sinh\tau)^{{\rm
i}\lambda}],
\end{equation}
is the relation
\begin{eqnarray}
\fl{{\cal Y}_2}^{l+\frac{1}{2}}_\lambda(\tau,\varphi) =
\frac{(-1)^l}{2{\rm i}\sqrt{2}\pi}\frac{1}{\prod_{r=1}^l\sqrt{r^2+\lambda^2}}
e^{{\rm
i}(l+\frac{1}{2})\varphi}\cosh^{l+\frac{1}{2}}\tau\nonumber\\
\fl\times\,\frac{d^l}{d(\sinh\tau)^l}\frac{1}{\sqrt{1+\sinh^2\tau}}
[(\sqrt{1+\sinh^2\tau}+\sinh\tau)^{{\rm
i}\lambda}-(\sqrt{1+\sinh^2\tau}-\sinh\tau)^{{\rm i}\lambda}].
\end{eqnarray}
Hence, we identify the hyperbolic counterpart ${\cal U}^l_\lambda$
of the Legendre functions which fulfils
\begin{equation}
{{\cal Y}_2}^{l+\frac{1}{2}}_\lambda(\tau,\varphi) = 
\frac{1}{2{\rm i}\sqrt{2}\pi}\frac{1}{\prod_{r=1}^l\sqrt{r^2+\lambda^2}}
e^{{\rm i}(l+\frac{1}{2})\varphi}{\cal U}^l_\lambda(\sinh\tau)
\end{equation}
such that
\begin{equation}
\fl{\cal U}^l_\lambda(x) =
(-1)^l(1+x^2)^{\frac{1}{2}(l+\frac{1}{2})}\frac{d^l}{dx^l}\frac{1}{\sqrt{1+x^2}}
[(\sqrt{1+x^2}+x)^{{\rm i}\lambda}-(\sqrt{1+x^2}-x)^{{\rm i}\lambda}].
\end{equation}
We remark that the form of the functions ${\cal T}^l_\lambda$ and
${\cal U}^l_\lambda$ suggests that they are somehow related to the
Chebyshev polynomials $T_n(x)$ and $U_n(x)$.
\section{Pseudospherical functions for continuous supplementary series}
We finally discuss the pseudospherical functions for the continuous
supplementary series.  We first introduce the parameter $\gamma$
such that
\begin{equation}
k=\gamma -\frac{1}{2},
\end{equation}
where in view of (2.14) $0<\gamma<\frac{1}{2}$.  By virtue of (2.13)
in this parametrization we can formally get the formulae for the
supplementary series from that obtained for the continuous
principal series by setting $\lambda=-{\rm i}\gamma$.  Therefore,
the even and odd spherical functions ${\hat{\cal Y}}^m_\gamma(\tau,\varphi)$
and ${\bar{\cal Y}}^m_\gamma(\tau,\varphi)$ for the continuous
supplementary series obtained from (5.1) and (5.2), respectively,
are given by
\begin{eqnarray}
\fl{\hat{\cal Y}}^m_\gamma(\tau,\varphi)=e^{{\rm i}m\varphi}{\hat
c}_{m\gamma}
\cosh^{-(\frac{1}{2}+\gamma)}\tau{}_2F_1[\hbox{$\scriptstyle{1\over 2}$}(m+
\hbox{$\scriptstyle{1\over 2}$}+\gamma),
\hbox{$\scriptstyle{1\over 2}$}(-m+\hbox{$\scriptstyle{1\over
2}$}+\gamma)
;\hbox{$\scriptstyle{1\over 2}$};{\rm tgh}^2\tau],\\
\fl{\bar{\cal Y}}^m_\gamma(\tau,\varphi)=e^{{\rm i}m\varphi}{\bar
c}_{m\gamma}
\cosh^{-(\frac{1}{2}+\gamma)}\tau{\rm tgh}\tau
{}_2F_1[\hbox{$\scriptstyle{1\over 2}$}
(m+\hbox{$\scriptstyle{3\over 2}$}+\gamma),
\hbox{$\scriptstyle{1\over 2}$}(-m+\hbox{$\scriptstyle{3\over
2}$}+\gamma)
;\hbox{$\scriptstyle{3\over 2}$};{\rm tgh}^2\tau].
\end{eqnarray}
Consider the even functions ${\hat{\cal Y}}^m_\gamma$.  An immediate
consequence of (5.5) is the relation
\begin{eqnarray}
&&\fl{}_2F_1[\hbox{$\scriptstyle{1\over 2}$}(m+
\hbox{$\scriptstyle{1\over 2}$}+\gamma),
\hbox{$\scriptstyle{1\over 2}$}(-m+\hbox{$\scriptstyle{1\over
2}$}+\gamma)
;\hbox{$\scriptstyle{1\over 2}$};{\rm tgh}^2\tau]\nonumber\\
&&\fl\quad{}={\hat A}_1\,{}_2F_1\left[\hbox{$\scriptstyle{1\over 2}$}(m+
\hbox{$\scriptstyle{1\over 2}$}+\gamma),
\hbox{$\scriptstyle{1\over 2}$}(-m+\hbox{$\scriptstyle{1\over
2}$}+\gamma)
;1+\gamma;\frac{1}{\cosh^2\tau}\right]\nonumber\\
&&\fl\qquad{}+{\hat A}_2\cosh^{2\gamma}\tau{}_2F_1\left[
\hbox{$\scriptstyle{1\over 2}$}(-m+
\hbox{$\scriptstyle{1\over 2}$}-\gamma),
\hbox{$\scriptstyle{1\over 2}$}(m+\hbox{$\scriptstyle{1\over
2}$}-\gamma)
;1-\gamma;\frac{1}{\cosh^2\tau}\right],
\end{eqnarray}
where
\numparts
\begin{eqnarray}
{\hat A}_1 &=& \frac{\sqrt{\pi}\Gamma(-\gamma)}{\Gamma[\hbox{$\scriptstyle{1\over 2}$}
(-m+\hbox{$\scriptstyle{1\over 2}$}-\gamma)]\Gamma[\hbox{$\scriptstyle{1\over
2}$}(m+\hbox{$\scriptstyle{1\over 2}$}-\gamma)]},\\
{\hat A}_2 &=& \frac{\sqrt{\pi}\Gamma(\gamma)}{\Gamma[
\hbox{$\scriptstyle{1\over 2}$}
(m+\hbox{$\scriptstyle{1\over 2}$}+\gamma)]\Gamma[\hbox{$\scriptstyle{1\over
2}$}(-m+\hbox{$\scriptstyle{1\over 2}$}+\gamma)]}.
\end{eqnarray}
\endnumparts
Notice that ${\hat A}_2(\gamma)={\hat A}_1(-\gamma)$.  Proceeding
analogously as with (5.1) we arrive at the following asymptotic form
of ${\hat {\cal Y}}^m_\gamma$:
\begin{equation}
\lim_{\tau\to\infty}{\hat{\cal Y}}^m_\gamma(\tau,\varphi)=e^{{\rm
i}m\varphi}{\hat
c}_{m\gamma}\cosh^{-\frac{1}{2}}\tau[B(\gamma)e^{-\gamma\tau}+
B(-\gamma)e^{\gamma\tau}],
\end{equation}
where $B(\gamma)B(-\gamma)={\hat A}_1{\hat A}_2$.  In opposition to
the functions (5.8) the norm of the functions (6.6) is divergent.
The same result holds true for the asymptotics of the odd functions
${\bar{\cal Y}}^m_\gamma(\tau,\varphi)$ as well.
\ack
This work was supported by the grant N N202 205738 from the National 
Centre for Research and Development.
\appendix
\section{}
We now collect the basic relations satisfied by the functions ${\cal
P}^m_k(x)$ given by (3.11) which are counterpart of the associated Legendre 
functions related to spherical harmonics.  Firstly, in view of (3.10), (2.20) 
and (2.21) ${\cal P}^m_k(x)$ satisfies the equation
\begin{equation}
\fl (1+x^2)\frac{d^2{\cal P}^m_k(x)}{dx^2}+2x\frac{d{\cal P}^m_k(x)}{dx}+\left[-k(k+1)+
\frac{m^2}{1+x^2}\right]{\cal P}^m_k(x) = 0.
\end{equation}
Further, an immediate consequence of differentiation of (3.11) is
\begin{equation}
(1+x^2)\frac{d{\cal P}^m_k(x)}{dx} = -\sqrt{1+x^2}{\cal P}^{m+1}_k(x)+mx{\cal P}^m_k(x).
\end{equation}
Eqs.\ (A.1) and (A.2) taken together yield
\begin{equation}
\fl {\cal P}^{m+2}_k(x)-2(m+1)\frac{x}{\sqrt{1+x^2}}{\cal P}^{m+1}_k(x)+(m-k)(m+k+1)
{\cal P}^m_k(x)=0.
\end{equation}
On reindexing $m\to m-1$ in (A.3) we can convert (A.2) into
\begin{equation}
\fl (1+x^2)\frac{d{\cal P}^m_k(x)}{dx} =
(m+k)(m-k-1)\sqrt{1+x^2}{\cal P}^{m-1}_k(x)-mx{\cal P}^m_k(x).
\end{equation}
The following identity can be obtained from (3.11) by reindexing
$m\to m-1$, $k\to k-1$ and using the Leibniz formula on $n$-th
derivative of product of functions
\begin{equation}
\fl 2k{\cal P}^m_{k-1}(x) =
\sqrt{1+x^2}{\cal P}^{m+1}_k(x)-2(m-k)x{\cal P}^m_k(x)+\sqrt{1+x^2}
(m-k)(m-k-1){\cal P}^{m-1}_k(x).
\end{equation}
Finally, reindexing $m\to m-1$ in (A.3) and combining the resulting
form of (A.3) with (A.5) we get
\begin{equation}
{\cal P}^m_{k-1}(x)-x{\cal P}^m_k(x) = (k-m+1)\sqrt{1+x^2}{\cal
P}^{m-1}_k(x).
\end{equation}
We point out the remarkable similarity of the structure of above
recurrences satisfied by the functions ${\cal P}^m_k(x)$ and
recursive relations for the Legendre functions $P^m_j(x)$ connected
with the spherical harmonics.  For example the counterpart of (A.2)
is of the form \cite{21}
\begin{equation}
(1-x^2)\frac{dP^m_j(x)}{dx} = -\sqrt{1-x^2}P^{m+1}_j(x)-mx
P^m_j(x),
\end{equation}
and the formula corresponding to (A.6) is given by 
\begin{equation}
P^m_{j-1}(x)-xP^m_j(x)=(j-m+1)\sqrt{1-x^2}P^{m-1}_j(x).
\end{equation}

\end{document}